\documentclass[journal,twoside,web]{ieeecolor}
\usepackage{tmi}
\usepackage{cite}
\usepackage{amsmath,amssymb,amsfonts}
\usepackage{algorithmic}
\usepackage{graphicx}
\usepackage{textcomp}

\usepackage[table]{xcolor}
\definecolor{coldgrey}{RGB}{128,128,105}
\usepackage[ruled,linesnumbered]{algorithm2e}
\usepackage{multicol}
\usepackage{multirow}
\usepackage{booktabs}

\def\BibTeX{{\rm B\kern-.05em{\sc i\kern-.025em b}\kern-.08em
    T\kern-.1667em\lower.7ex\hbox{E}\kern-.125emX}}
\begin{document}
\title{Federated Pseudo Modality Generation for
Incomplete Multi-Modal MRI Reconstruction}
\author{Yunlu Yan, Chun-Mei Feng, Yuexiang Li, Rick Siow Mong Goh, Lei Zhu
\thanks{Yunlu Yan is with The Hong Kong University of Science and Technology (Guangzhou), Nansha, Guangzhou, 511442, Guangdong, China.~(Email: yyan538@connect.hkust-gz.edu.cn)}
\thanks{Chun-Mei Feng and Rick Siow Mong Goh are with the
Institute of High Performance Computing, A*STAR, Singapore, 138632,    China.~(Email: strawberry.feng0304@gmail.com,  gohsm@ihpc.a-star.edu.sg).}
\thanks{Yuexiang Li is with Guangxi Medical University, Nanning, 530021, China.~(Email: leeyuexiang@163.com)}
\thanks{Lei Zhu is with The Hong Kong University of Science and Technology (Guangzhou), Nansha, Guangzhou, 511442, Guangdong, China and The Hong Kong University of Science and Technology, Hong Kong SAR, 999077, China.~(Email: leizhu@ust.hk)}
\thanks{Corresponding authors: \textit{Lei Zhu.}}
\thanks{Y.Yan and C.Feng contributed equally to this work.}
}

\maketitle

\begin{abstract}
While multi-modal learning has been widely used for MRI reconstruction, it relies on paired multi-modal data which is difficult to acquire in real clinical scenarios. Especially in the federated setting, the common situation is that several medical institutions only have single-modal data, termed the modality missing issue. Therefore, it is infeasible to deploy a standard federated learning framework in such conditions.
In this paper, we propose a novel communication-efficient federated learning framework, namely Fed-PMG, to address the missing modality challenge in federated multi-modal MRI reconstruction.
Specifically, we utilize a pseudo modality generation mechanism to recover the missing modality for each single-modal client by sharing the distribution information of the amplitude spectrum in frequency space. 
However, the step of sharing the original amplitude spectrum leads to heavy communication costs. To reduce the communication cost, we introduce a clustering scheme to project the set of amplitude spectrum into finite cluster centroids, and share them among the clients. With such an elaborate design, our approach can effectively complete the missing modality within an acceptable communication cost. Extensive experiments demonstrate that our proposed method can attain similar performance with the ideal scenario, i.e., all clients have the full set of modalities. The source code will be released.
\end{abstract}


\section{Introduction}

Magnetic resonance imagining (MRI) reconstruction is an effective way to accelerate the imaging of MRI. Recently, the models utilizing the paired multi-modal data~\cite{xiang2018deep,feng2022multi,xuan2022multi} have been proved to achieve a progressive reconstruction performance, \emph{i.e.,} using an acquired full-sampled auxiliary modality to guide the reconstruction of under-sampled target modality. However, the existing methods are data-driven and need a lot of paired multi-modal data for training, which is difficult for a single hospital to satisfy. Moreover, due to the privacy issue, it’s infeasible to share highly sensitive patient data in multiple institutions.
\begin{figure}[!t]
    \centering
    \includegraphics[width=1.0\linewidth]{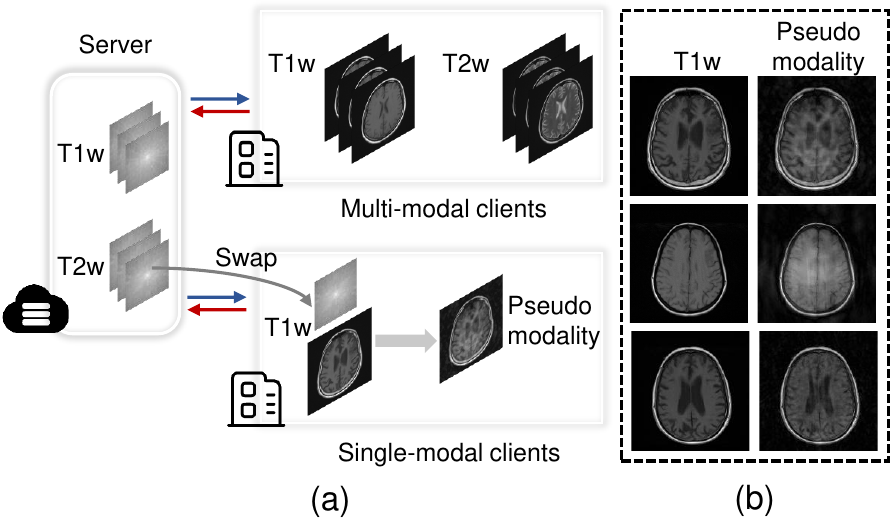}
    \caption{(a) The task setting of our incomplete multi-modal FL where only part of the clients have the full set of modalities, which aims to learn a global multi-modal reconstruction model under the modality missing problem. To this end, we swap the shared amplitude spectrum of the missing modality with its original amplitude spectrum at the single-modal client. (b) The examples of original data and generated pseudo modality. }
    \label{fig:intro}
\end{figure}

Federated learning (FL) is a distributed learning paradigm that allows multiple institutions to collaboratively train a model without privacy disclosure, which can address the above challenge. Several recent studies~\cite{guo2021multi,feng2022specificity} have introduced FL into
MRI reconstruction to mitigate the data dependence of deep learning-based methods. Nevertheless, we surprisingly find that existing FL-based reconstruction methods are built upon the single-modal reconstruction task, which impedes the wide applicability of FL in clinical deployment. In terms of multi-modal reconstruction in a federated setting, an ideal scenario is that all clients have paired multi-modal data. However, due to the expensive costs of collecting MR data, it’s impractical that every hospital has a full set of modalities in real clinical scenarios. For some small hospitals, some modalities may be missing, which makes it impossible for them to participate in FL. Thus, the standard
FL framework~\cite{mcmahan2017communication} is unable to conduct in such a scenario.

Modality missing is a very practical problem, which has been considered in different tasks, \emph{e.g.}, disease diagnosis~\cite{liu2021incomplete,pan2019disease}, and segmentation~\cite{chen2019robust}, but it is not fully explored in FL paradigm. The distributed data of multiple clients pose new challenges for modality missing. To this end, we study a novel problem setting of \textit{incomplete multi-modal federated learning}, namely IMFL, for MRI reconstruction, where several clients have paired multi-modal data while the rest of the clients only have single-modal data, as shown in Fig.~\ref{fig:intro} (a). There are two primary challenges in this problem setting: (\expandafter{\romannumeral1}) The unbalanced modality of different clients hinders the global deployment of FL. (\expandafter{\romannumeral2}) 
The data in multiple institutions is sampled by the different scanners which will cause the domain shift even in the ideal scenario. A feasible solution for IMFL is to complete the missing modalities, which let every single-modal client trains a local multi-modal reconstruction model such that we can further obtain a globally consistent model.

In this work, we propose a communication-efficient \textbf{Fed}erated \textbf{P}seudo \textbf{M}odality \textbf{G}eneration mechanism (Fed-PMG), to address the above two problems. Our insight is to generate the missing modality by utilizing inherent knowledge in frequency space. As for the frequency space of an image, the amplitude spectrum can capture low-level distribution information  (\emph{i.e.,} style) while the phase spectrum can capture high-level content information (\emph{i.e.,} structure)~\cite{liu2021feddg,yang2020fda}. Meanwhile, sharing the amplitude spectrum can address the domain shift problem to some extent~\cite{yang2020fda,jiang2022harmofl,liu2021feddg}. The idea is motivated by an observation that combining the phase spectrum of modality \textit{T1w} with the  amplitude spectrum of modality \textit{T2w} can generate a new image, called pseudo modality that has the same structure as modality \textit{T1w} and keeps the style of modality \textit{T2w}, as shown in Fig.~\ref{fig:intro} (b).
In addition, it will bring vast communication costs if we shared the amplitude spectrum of all clients. Thus, before sending the amplitude spectrum from client to server, we exploit a clustering strategy to transform the amplitude spectrum to the cluster centroid that can capture the distribution information. Compared with sharing the original amplitude spectrum, our approach requires less communication cost. Then, we use the shared centroid to complete the missing modality of the single-modal clients, which enables each client to train a multi-modal MRI reconstruction model. Moreover, the centroid contains the distribution information, which can mitigate the domain shift problem. In a nutshell, our main contributions can be summarized as follows:
\begin{itemize}
	\setlength{\itemsep}{2pt}
	\setlength{\parsep}{-2pt}
	\setlength{\parskip}{0pt}
	\setlength{\leftmargin}{-15pt}
\item {We uncover a practical problem in the real-world application, \emph{i.e.,} learning a multi-modal model in the modality missing setting. To the best of our knowledge, this is the first work studying incomplete multi-modal learning for MRI reconstruction in the federated setting.}

\item {We propose a communication-efficient method, Fed-PMG, that utilizes a novel federated pseudo modality generation mechanism to complete the modality of the single-modal clients.}

\item{Extensive experiments demonstrate that our method can effectively complete the missing modality and achieve similar performance with the ideal scenario. With the proposed clustering strategy, the communication cost greatly drops by $97.5\%$.}

\end{itemize}

\section{Related Work}
\noindent\textbf{Multi-modal MRI Reconstruction.} MRI Reconstruction~\cite{knoll2020deep} aims to accelerate the imaging of MR via reconstructing the under-sampled measure of frequency space to high-quality MR images, which is essentially an ill-posed inverse problem. To solve this problem, the traditional single-modal~\cite{wang2016accelerating,quan2018compressed,sriram2020end} methods only learned the features from single-modal MR images to reconstruct the under-sampled images. Recently, with the development of multi-modal learning, it received a lot of attention in medical imaging including MRI reconstruction~\cite{sun2019deep,chen2019robust,wang2021transbts}. Multi-modal MRI reconstruction leverages the paired multiple modalities to reconstruct high-quality MR images, which has been demonstrated that can achieve better performance than single-modal methods. The existing methods can be divided into two categories: joint reconstruction~\cite{majumdar2011joint,sun2019deep,liu2020multi} and auxiliary reconstruction ~\cite{xuan2022multi,feng2022multi,xiang2018deep,lyu2022dudocaf}. The former reconstruct multiple under-sampled modalities at the same time, in contrast, the latter use the auxiliary modality that is full-sampled to guide the reconstruction of the under-sampled target modality. Compared with the joint reconstruction methods, the auxiliary reconstruction methods are more reliable and have better reconstruction performance, which is explored in this paper. To utilize the complementary information of full-sampled auxiliary modality, Xiang \textit{et al.}~\cite{xiang2018deep} directly merged two modalities as input and then fed it into the reconstruction network. For better interaction, Feng \textit{et al.}~\cite{feng2022multi} introduced a novel cross-transformer network to repeatedly fuse the features of two modalities. Differently, Xuan \textit{et al.}~\cite{xuan2022multi} considered the additional problem that the auxiliary modality is misaligned. Although the previous works have achieved success, they explore the multi-modal methods in a centralized training paradigm. There still exists unsolved multi-modal problems in the discrete multi-client scenario, a more complex surrounding.
\begin{figure*}[!t]
    \centering
    \includegraphics[width=1.0\textwidth]{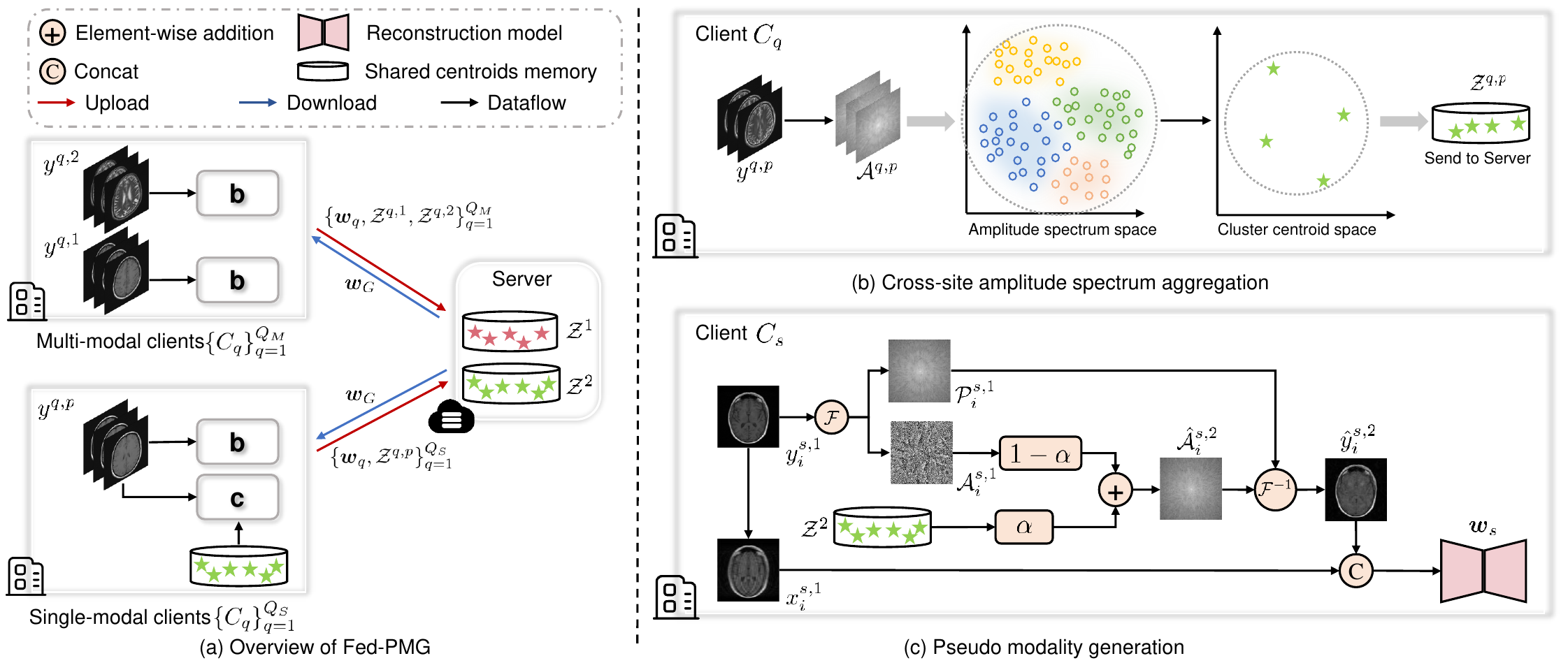}
    \caption{(a) An overview of the proposed federated pseudo modality generation framework, including two fundamental modules: (b) Cross-site amplitude spectrum aggregation. Every client aggregates the amplitude spectrum of all images and projects them to cluster centroid, which is further sent to the server. (c) Pseudo modality generation. The single-modal client, \emph{e.g.},  $C_{s}$, utilizes the shared centroid to generate the missing modality. After the above two stages, we can train a multi-modal reconstruction model.}  \label{fig:overview}
\end{figure*}

\noindent\textbf{Federated Learning.} FL provides a general solution for distributed training, which receives wide attention~\cite{li2021survey}. One of the standard architectures, FedAvg~\cite{mcmahan2017communication}, learned an optimal global model by averaging the updated model parameters of each client. Due to its effectiveness and communication efficiency, FedAvg has been the most popular method. However, due to the client-drift, the performance of FedAvg will decrease on heterogeneous data~\cite{li2019convergence, karimireddy2020scaffold}. To solve this problem, they proposed variants based on the FedAvg~\cite{wang2020tackling,li2020federated, li2021fedbn}. The significant benefit of FL is that it can leverage the data of multiple clients as long preserving privacy, which is greatly
 suitable for application in the medical imaging field~\cite{dou2021federated,liu2021feddg,rieke2020future,li2019privacy,liu2021semifed}, a privacy-sensitive and data-constrained scenario. In terms of MRI reconstruction, Guo \textit{et al.}~\cite{guo2021multi} first introduced FL into MRI reconstruction, namely FL-MRCM, to mitigate the data dependence of deep learning-based method and address the data heterogeneity of clients. FL-MRCM aligned the feature distributions of multiple source sites to a target site by adversarial learning. Based on the setting of FL-MRCM, Feng \textit{et al.}~\cite{feng2022specificity} proposed a novel FL framework based on personalized FL to address the domain shift of different clients by learning the heterogeneous information of each client. However, the existing FL-based MRI reconstruction methods are based on the single-modal paradigm, which seeks to address the data heterogeneity from multiple data domains. In contrast, our goal is to train a global multi-modal reconstruction model with only incomplete multi-modal data.

\noindent\textbf{Frequency Space Decomposition.} 
Frequency space decomposition is an effective technique for transforming domain-specific knowledge, as it leverages the amplitude spectrum's rich style information. This has proven valuable in various applications, such as domain adaption~\cite{yang2020fda} and domain generation~\cite{yao2022enhancing}.  In Federated Learning, FedDG~\cite{liu2021feddg} shared the distribution information across clients and generated the new domain by frequency space interpolation, thereby improving the generation of the model. Besides, Harmofl~\cite{jiang2022harmofl}  introduced this ingenious mechanism to address the domain shift between different clients. While frequency space decomposition has shown promising results in various applications, its potential in addressing federated incomplete multi-modal learning remains largely unexplored. Additionally, directly sharing the amplitude spectrum can lead to substantial communication costs. This paper aims to fill this research gap by focusing on applying frequency space decomposition techniques to tackle the missing modality challenge. To enhance communication efficiency, the paper proposes an innovative utilization of the clustering method.
 
\section{Preliminaries}
\label{sec:pre}
In the IMFL setting, as shown in Fig.~\ref{fig:overview} (a), there are $Q$ clients $\{C_{q}\}_{q=1}^Q$, where only $Q_{M}$ clients have a paired multi-modal dataset and the rest $Q_{S}$ clients have a single-modal dataset, which contains one modality of the full set, \emph{i.e.,} $Q = Q_{M}+Q_{S}$. For the multi-modal dataset $D_{M} \in \{D_{q,M}\}_{q=1}^{Q_{M}}$, it has $n_{M}$ image pairs $ \{(y_{i}^{1}, y_{i}^{2})\}_{i=1}^{n_{M}}$. In constrat, the single-modal dataset  $D_{S} \in \{D_{q,S}\}_{q=1}^{Q_{S}}$ has $n_{S}$ images $\{y_{i}^{p}\}_{i=1}^{n_{S}}$, where $p \in \{1,2\}$. Supposing that $y^1$ is the target modality and $y^2$ is the auxiliary modality, our goal is to train a global optimal multi-modal reconstruction model under the above missing modality setting.

\noindent\textbf{Multi-modal FL for MRI Reconstruction.} Let $\mathcal{M}$ denote the mask matrix that represents the sample point of frequency space. The under-sampled image $x$ can be obtained from the ground truth image $y$ via:
\begin{equation} 
x=\mathcal{F}^{-1}\left(\mathcal{M} \odot \mathcal{F}(y)+\epsilon\right),
\label{eq:ds}
\end{equation}
where $\epsilon$ is the measurement noise, $\mathcal{F}$ is the multidimensional Fourier transform, and $\mathcal{F}^{-1}$ is the inverse operation of multidimensional Fourier transform. One multi-modal MRI reconstruction paradigm is to utilize the easily acquired modality to help the reconstruction of other modality, which needs long-time scanning. With multi-modal data, the simplest way to obtain the multi-modal information is channel fusing~\cite{xiang2018deep} that directly concatenates the two modalities as input before feeding them into the model, which can be written as:
\begin{equation}
\mathcal{L}_{recon} = \| f(x^{1} \biguplus y^{2}; \boldsymbol{w}) - y^{1}\|_{1},
\label{eq:mrecon}
\end{equation}
where $(x^{1}, y^{1})$ is a pair of the target modality, $y^{2}$ is the ground truth of auxiliary modality, $\biguplus$ is the concat operation, $\| \cdot \|_1$ is the $\mathcal{L}_1$ loss, $f$ is the multi-modal MRI reconstruction model and $\boldsymbol{w}$ is the parameters of the model. If all clients have complete multi-modal data, we can train a local multi-modal model at each client.
Considering the standard FL framework~\cite{mcmahan2017communication}, the global loss can be formulated as:
\begin{equation}
\mathcal{L} = \sum_{q=1}^{Q} p_{q} \mathcal{L}_{recon}, \quad \text{where} \    p_{q}=\frac{n_{q}}{\sum_{i=1}^{Q}n_{i}}.
\label{eq:1}
\end{equation}
The updated model parameters of the clients will be transferred to the server, and then the server averages the parameters of all local models to get the global model, which can be written as:
\begin{equation}
    \boldsymbol{w}_{G} = \sum_{q=1}^{Q} p_{q}\boldsymbol{w}_{q}. \label{eq:avg}
\end{equation}
However, some clients only have the single-modal dataset in our setting. It's infeasible that they train a multi-modal model in a general way. A naive way is to regard the multi-modal clients as single-modal clients and train a single-modal model with mixed data. However, such a paradigm ignores the precious information of multi-modal data. In addition, it is unable to update the global model due to the model heterogeneity if we learn the different models for different types of clients. To address these issues, we introduce the Fed-PMG framework. In the subsequent section, we will provide a detailed and thorough description of our approach.

\section{Fed-PMG}
As stated earlier, we can utilize the knowledge of the amplitude and phase spectrum in frequency space to generate the missing modality. However, in the IMF setting, some clients only have the single modality, which can not access the amplitude spectrum of other modalities. To overcome this issue, we propose a communication-efficient federated pseudo modality generation scheme, as shown in Fig.~\ref{fig:overview}, including two key modules, \emph{i.e.,} cross-site amplitude spectrum aggregation and pseudo modality generation.

\subsection{Cross-site Amplitude Spectrum Aggregation}
Fig.~\ref{fig:overview} (b) depicts the details of cross-site amplitude spectrum aggregation module. To be specific, given a full-sampled image $y_{i}^{q,p}$ from the client $C_{q}$, we first transform the image from the image domain to the frequency domain $\mathcal{K}_{i}^{q,p}$ via multidimensional Fourier transform:
\begin{equation}
\mathcal{K}_{i}^{q,p} = \mathcal{F}(y_{i}^{q,p}), \quad \text{and} \ p=1,2.
\label{eq:4}
\end{equation}
Then, we can obtain the amplitude spectrum $\mathcal{A}_{i}^{q,p}$ and phase spectrum $\mathcal{P}_{i}^{q,p}$ from $\mathcal{K}_{i}^{q,p}$ (via $\mathcal{A}_{i}^{q,p} \circ \mathcal{P}_{i}^{q,p} = \mathcal{K}_{i}^{q,p}$), which can capture the style information and structure information, respectively. Before the training of the client, the above procedure will be repeated until we get the amplitude spectrum of all images $\mathcal{A}^{q,p} = \{\mathcal{A}_{i}^{q,p}\}_{i=1}^{n_{q}}$.

A naive way to share the amplitude spectrum is to send $\mathcal{A}^{q,p}$ to the server directly. However, such an approach has two fatal flaws. \emph{First}, the cost of storage and communication is coupled with the data quantity of the client. Thus, it needs a mass of storage and communication costs, while the amount of client data is large. \emph{Second}, there still exists some redundant information in the amplitude spectrum of different MR images, when the amplitude spectrum from all clients are gathered in the server. To this end, for each modality, we use the clustering strategy to group the amplitude spectrum into clusters and further obtain the centroid of clusters, which can be described as follows:
\begin{equation}
\{\mathcal{Z}_{1}^{q,p},\mathcal{Z}_{2}^{q,p}, \ldots, \mathcal{Z}_{K}^{q,p}\} = cluster(\mathcal{A}^{q,p}),
\label{eq:clu}
\end{equation}
where $\mathcal{Z}$ is the the cluster centroid of amplitude spectrum, and $K$ is the number of cluster centroid. It is noted that the number of cluster centroid $K$ is far smaller than the number of the original amplitude spectrum, which can greatly decrease the cost of storage and communication. Finally, the cluster centroid capturing the distribution information of the dataset will be sent to the server, while the multi-modal clients will send the cluster centroid of two modalities. The server then gathers the centroid of all clients $\{\mathcal{Z}^{1}, \mathcal{Z}^{2}\}$ and saves them into the memory according to their modality. Notably, the aggregation of the amplitude spectrum is conducted only once before training.

\noindent{\textbf{Privacy Preserving.}} As suggested by previous works~\cite{liu2021feddg, jiang2022harmofl}, the amplitude spectrum (see Fig.~\ref{fig:overview}) only contains the distribution information of the MR image, which makes it infeasible to invert the original image. Therefore, our method is privacy-secure.

\begin{algorithm}[!t]
\caption{Fed-PMG}
\label{alg:immfl}
\KwIn{$Q_{M}$ multi-modal datasets $\{D_{q,M}\}_{q=1}^{Q_{M}}$,
$Q_{S}$ single-modal datasets $\{D_{q,S}\}_{q=1}^{Q_{S}}$}
\KwOut{Pseudo modality}
\textbf{Amplitude Spectrum Aggregation:}\\
\For{client $q=1,2,...,Q_M+Q_S$ \textbf{parallelly}}{
    \For{modality $p=1,2$}{
        Gather the amplitude spectrum $\mathcal{A}^{q,p}$\;
        $\mathcal{Z}^{q,p}$ $\leftarrow$ $cluster(\mathcal{A}^{q,p})$\;
        Send $\mathcal{Z}^{q,p}$ to the server\;
    }
    
}
\textbf{Server Execute:}\\
Gather the centroid from all clients\;
Build up the centroid memory $\{\mathcal{Z}^{1},\mathcal{Z}^{2}\}$ \;
\textbf{Pseudo Modality Generation:}\\
\For{client $s=1,2,...,Q_{S}$ \textbf{parallelly}}{

    $h,p \in \{1, 2\}$ and $h \neq p$ \;
    Receive the centorids from the server $\mathcal{Z}^{h}$\;

    \For{batch $b=(x^{s,p},y^{s,p})$ from $\mathcal{D}_{s,S}$}{
        $\mathcal{Z}^{h}_j$ $\leftarrow$ $Random(\mathcal{Z}^{h})$\;
           $\hat{\mathcal{A}}^{s,h}_{i} $ $\leftarrow$  $(1 -\alpha) \mathcal{A}_{i}^{s,p} + \alpha \mathcal{Z}_{j}^{h}$\;
    $\hat{y}_{i}^{s,h}$ $\leftarrow$ $\mathcal{F}^{-1}(\hat{\mathcal{A}}^{s,h}_{i}\circ \mathcal{P}^{s,p}_{i})$\;
    }
}
\end{algorithm}

\subsection{Pseudo Modality Generation}
The clients containing multi-modal data can train a multi-modal model via Eq.~\eqref{eq:mrecon}, while it is unable to apply to single-modal clients. Therefore, we propose a pseudo modality generation module to complement the missing modality by sharing the cluster centroid instead of the original amplitude spectrum in terms of the single-modal clients, as shown in Fig.~\ref{fig:overview}(c). In detail, we first share the cluster centroid memory $\mathcal{Z}^{2}$ to client $C_s$, which only has the $y^1$ modality. Then, we randomly select a centroid $\mathcal{Z}_{j}^{2}$ from centroid memory  $\mathcal{Z}^{2}$ and combine it with the amplitude spectrum of client $C_s$ by weighting as follows:
\begin{equation}
    \hat{\mathcal{A}}^{s,2}_{i} = (1-\alpha) \mathcal{A}_{i}^{s,1} + \alpha \mathcal{Z}_{j}^{2},
\end{equation}
where $\alpha$ is the parameter to control the contribution of the centroid, and $\hat{\mathcal{A}}^{s,2}_{i}$ is the new amplitude spectrum of pseudo modality that is missing in client $C_{s}$. 

Then, we combine $\hat{\mathcal{A}}^{s,2}_{i}$ with the original phase spectrum $\mathcal{P}^{s,1}_{i}$ and feed them into the inverse multidimensional Fourier transform function to generate the image of pseudo modality $\hat{y}^{s,2}_{i}$. The procedure can be described as:
\begin{equation}
\hat{y}_{i}^{s,2} = \mathcal{F}^{-1}(\hat{\mathcal{A}}^{s,2}_{i}\circ \mathcal{P}^{s,1}_{i}).
\label{eq:5}
\end{equation}
Similarly, we can generate $\hat{y}^{1}$ modality for the single-modal clients, which only have $y^{2}$ modality in the same manner. The subtle difference is that we use the $y^{2}$ modality to guide the reconstruction of $\hat{y}^{1}$ modality.

\subsection{Local Training}
With the generated pseudo modality $\hat{y}_{i}^{s,h}, h\in\{1, 2\}$, we can use the standard FL framework~\cite{mcmahan2017communication} to train a global multi-modal model. As stated in \S\ref{sec:pre}, the local objective of single-modal clients can be formulated as:
\begin{equation}
\mathcal{L}_{recon} = \begin{cases} \| f(\hat{x}^{s,h}_{i} \biguplus y^{s,p}_{i}; \boldsymbol{w}) - \hat{y}^{s,h}_{i}\|_{1}, & h=1, h \neq p  \\
\| f(x^{s,p}_{i} \biguplus \hat{y}^{s,h}_{i}; \boldsymbol{w}) - y^{s,p}_{i}\|_{1}, & h=2, h \neq p 
\end{cases}.
\label{eq:mrpse}
\end{equation}
Additionally, the local model of multi-modal clients can be optimized via Eq.~\eqref{eq:mrecon}. After the local training, the clients send the model parameters to the server and we can update the global model via Eq.~\eqref{eq:avg}. The detailed information of our method is shown in Alg.~\ref{alg:immfl}.

\begin{table*}[!t]
	\centering
		\caption{\small Quantitative comparison of all methods over two modalities, where 
		\textit{UB} indicates the upper-bound. The best and second-best results of FL methods are marked in red and blue, respectively. }
		\small
		\resizebox{1.0\textwidth}{!}{
		\setlength\tabcolsep{4pt}
		\renewcommand\arraystretch{1.5}{
		\begin{tabular}{r||cccc|cccc|cccc|cccc}
        \hline
		\multirow{3}{*}{Method} &\multicolumn{4}{c|}{\textbf{fastMRI 1.5T}}
		 &\multicolumn{4}{c|}{\textbf{fastMRI 3T}}
		 &\multicolumn{4}{c|}{\textbf{uMR}}
		 &\multicolumn{4}{c}{\textbf{Average}} \\
		 \cline{2-17}
		 &\multicolumn{2}{c}{\textbf{T1w}}
		 &\multicolumn{2}{c|}{\textbf{T2w}}
		 &\multicolumn{2}{c}{\textbf{T1w}}
		 &\multicolumn{2}{c|}{\textbf{T2w}}
		 &\multicolumn{2}{c}{\textbf{T1w}}
		 &\multicolumn{2}{c|}{\textbf{T2w}}
		 &\multicolumn{2}{c}{\textbf{T1w}}
		 &\multicolumn{2}{c}{\textbf{T2w}}\\
		 &PSNR&SSIM
		 &PSNR&SSIM
		 		 &PSNR&SSIM
		 &PSNR&SSIM
		 		 &PSNR&SSIM
		 &PSNR&SSIM
		 		 &PSNR&SSIM
		 &PSNR&SSIM\\
		 
		 \hline
		  \hline
		 &\multicolumn{16}{c}{\textbf{Random 4$\times$}}\\
			\hline
		 \textit{Group}&32.9&0.839&31.5&0.917
		 &32.5&0.762&32.6&0.930
		 &30.4&0.859&31.3&0.916
		 &32.0&0.820 & 31.8&0.921 \\
		 
		 \textit{Mixup}&31.6&0.789&29.7&0.720
		 &30.6&0.702&30.8&0.749
		 &30.2&0.820&30.5&0.784
		 &31.1&0.770 & 30.3&0.751 \\
		 
		 \textit{Ideal}&\color{blue}{36.3}&\color{red}{0.951} &\color{blue}{32.9}&\color{blue}{0.937}
		 &\color{red}{35.2}&\color{red}{0.955} &\color{blue}{33.2}&\color{blue}{0.944}
		 &\color{blue}{33.4}&\color{red}{0.944} & \color{red}{33.0}&\color{blue}{0.946}
		 &\color{blue}{35.0}&\color{red}{0.950} & \color{blue}{33.1}&\color{blue}{0.942}\\
		 
		  		 \textit{Gather(UB)} &  37.0&0.955 & 33.5&0.944
 		 &37.2&0.962&34.1&0.951
 		 &35.8&0.964&34.2&0.958
 		 &36.7&0.960 & 33.9&0.951 \\

		 \textbf{Fed-PMG}&\color{red}{36.4}&\color{blue}{0.950}&\color{red}{33.0}&\color{red}{0.938}
		 &\color{blue}{35.1}&\color{blue}{0.952}&\color{red}{33.3}&\color{red}{0.944}
		 &\color{red}{33.5}&\color{blue}{0.940}&\color{blue}{33.0}&\color{red}{0.946}
		 &\color{red}{35.0}&\color{blue}{0.947}&\color{red}{33.1}&\color{red}{0.943} \\
		 \hline\hline
		  &\multicolumn{16}{c}{\textbf{ Equispaced 4$\times$}}\\
		  \hline
		  
		  \textit{Group}&35.9&0.937&32.9&0.930
		  &36.9&0.952&34.2&0.941
		  &33.3&0.928&32.5&0.930
		  &35.4&0.939&33.2&0.934 \\
		  
		  \textit{Mixup}&32.3&0.794&30.3&0.724
		  &31.0&0.709&31.5&0.751
		  &31.3&0.831&31.3&0.784
		  &31.6&0.776&31.0&0.753 \\

		  \textit{Ideal}&\color{blue}{37.8}&\color{red}{0.958} &\color{blue}{34.4}&\color{blue}{0.947}
		 &\color{blue}{37.5}&\color{red}{0.963} &\color{blue}{34.8}&\color{blue}{0.951}
		 &\color{blue}{35.0}&\color{red}{0.954} & \color{blue}{34.4}&\color{blue}{0.951}
		 &\color{blue}{36.8}&\color{red}{0.959} & \color{blue}{34.5}&\color{blue}{0.950}\\
		 
 		 \textit{Gather(UB)} &  38.5&0.962 & 34.9&0.952
 		 &38.5&0.967&35.7&0.958
 		 &37.2&0.969&35.5&0.963
 		 &38.1&0.966 & 35.4&0.958 \\

		  \textbf{Fed-PMG}&\color{red}{38.0}&\color{blue}{0.957}&\color{red}{34.4}&\color{red}{0.948}
		 &\color{red}{37.5}&\color{blue}{0.962}&\color{red}{34.9}&\color{red}{0.953}
		 &\color{red}{35.1}&\color{blue}{0.951}&\color{red}{34.5}&\color{red}{0.954}
		 &\color{red}{36.9}&\color{blue}{0.957}&\color{red}{34.6}&\color{red}{0.951} \\
		 
		\hline
		\end{tabular}
		}}
	\label{tab1}
\end{table*}

\begin{figure*}[t]
\centering
  \includegraphics[width=1.0\textwidth]{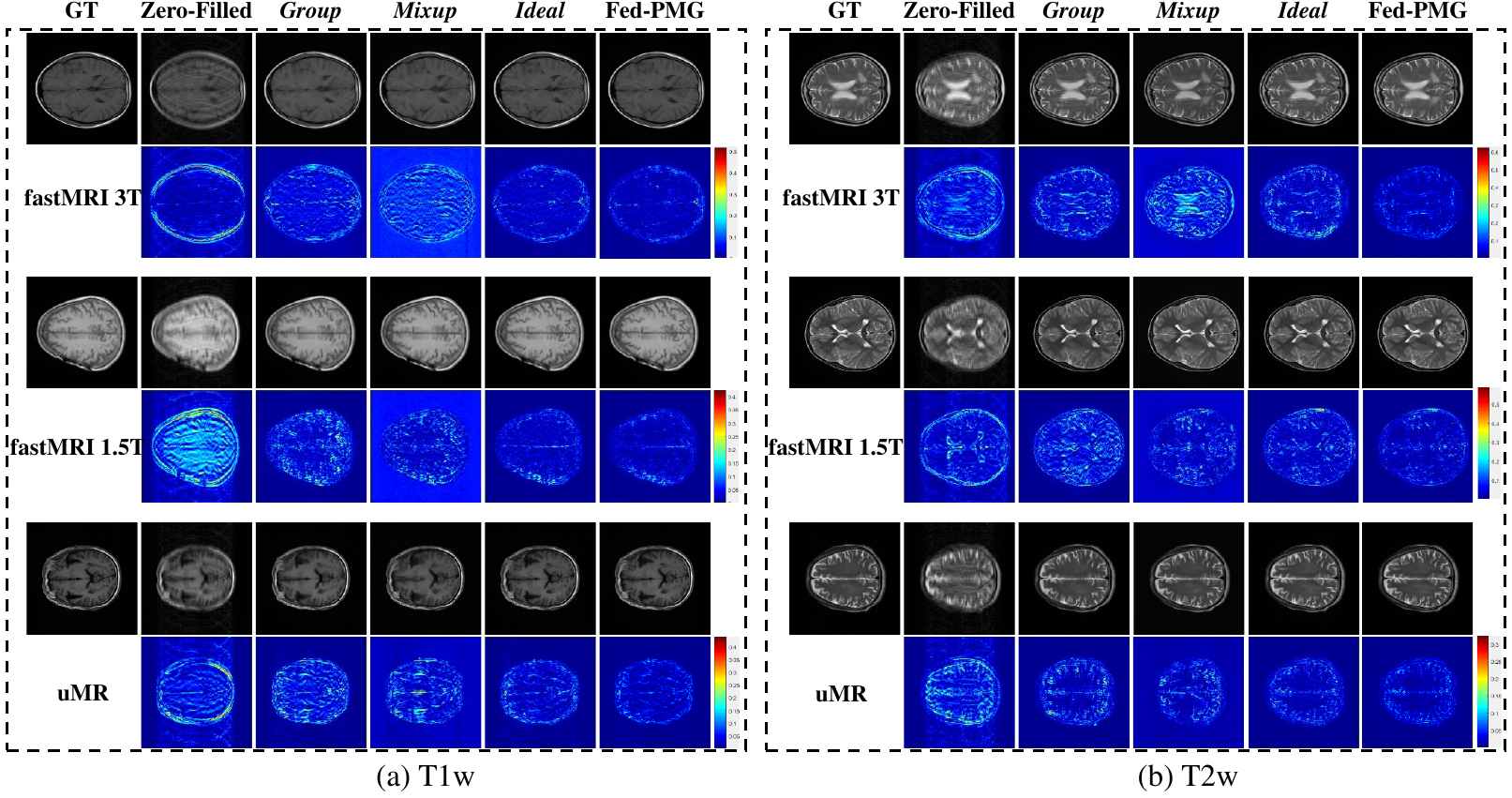}
  \caption{Reconstructed MR images and error maps with different methods on the fastMRI~\cite{zbontar2018fastmri} dataset. We present the visualization results from different clients and modalities: (a) T1w (b) T2w.}
  \label{tab:errormap}
\end{figure*}

\section{Experiments}
\subsection{Experiment setting}
\noindent\textbf{Datasets.} We evaluate the effectiveness of our method on the public fastMRI brain~\cite{zbontar2018fastmri} dataset and a clinical brain uMR dataset. The size of all images is $320 \times 320 $ pixels. To be specific, we select the paired T1-weighted (T1w) modality and T2-weighted (T2w) modality from the fastMRI dataset with 200 subjects. Each subject has 16 T1w and 16 T2w slices. 
All subjects are full-sampled brain MR images obtained on 3 or 1.5 Tesla magnets, where 109 subjects are sampled by 3 Tesla magnets and the rest of the subjects are sampled by 1.5 Tesla magnets. Besides, the uMR~\cite{feng2022specificity} dataset is collected by United  Imaging  Healthcare with UMR 790  3T scanner and contains 42 subjects, where each subject has 20 T1w and 20 T2w slices.

\noindent\textbf{Clients Setting.} To simulate the data from multiple centres, we first split the fastMRI dataset into two sub-datasets according to the type of scanner. 
By taking a dataset as a client, we can obtain three clients: \textbf{fastMRI 3T}, \textbf{fastMRI 1.5T}, and \textbf{uMR}. For each dataset, we randomly divide it into a training set and a testing set with a ratio of 7:3. 
Here, the fastMRI 3T is a multi-modal client and utilizes both the T1w modality and the T2w modality, while fastMRI 1.5T and uMR are the single-modal client.
The fastMRI 1.5T client uses only the T1w modality, and the uMR client utilizes only the T2w modality.
Noted that the above setting is randomly set, which does not affect the reliability of the experiment.

\noindent\textbf{Implementation Details.} We train all methods on one NVIDIA RTX 3080 GPU with 10 GB memory. All code is implemented by Pytorch and the batch size is 8. The local training is optimized by Adam optimizer with a learning rate of 0.0001. 
Moreover, the number of communication rounds is 50 with 5 local epochs per round. 
The above setting is used in all clients. 
Moreover, all methods are conducted in the same environment and have converged at the end of training. We use K-Means~\cite{hartigan1979algorithm} as the cluster method to project the amplitude spectrum into cluster centroid, and the two hyperparameters $K$ and $\alpha$ are empirically set to $50$ and $0.09$, respectively.

\noindent\textbf{Baselines:} To verify the effectiveness of our method, we compare it with the following baselines:
\begin{itemize}
	\setlength{\itemsep}{3pt}
	\setlength{\parsep}{-2pt}
	\setlength{\parskip}{-0pt}
	\setlength{\leftmargin}{-15pt}
	\item \textbf{\textit{Ideal}}: We introduce the unused real modality to single-modal clients. Thus, all clients have paired multi-modal data, which is the 
    ideal situation for multi-modal federated learning.
    \item \textbf{\textit{Mixup}}: We treat each client as a single-modal client and train a global single-modal reconstruction model via FedAvg~\cite{mcmahan2017communication}. Therefore, for multi-modal clients, \textit{Mixup} feeds two modalities into the model without making a distinction.
    \item \textbf{\textit{Group}}: Based on Mm-FedAvg~\cite{zhao2022multimodal}, we let clients train the single-modal model for two modalities (i.e, T1w and T2w), and then average model parameters trained from the same modality in the server-site by grouping aggregation.
    
    \item \textbf{\textit{Gather}}: In the ideal scenario, we gather the trainset of three clients in a client and then train a multi-modal reconstruction model, which is the theoretical upper bound\cite{guo2021multi} of multi-modal FL.
    
\end{itemize}
We employ the U-Net model to conduct reconstruction tasks. And especially, the \textit{Ideal}, \textit{Gather} and our method use channel fusing to learn a multi-modal model (\emph{i.e.}, Eq.~\eqref{eq:mrecon}). For all methods, we use FedAvg~\cite{mcmahan2017communication} to train a global model.

\subsection{Results Evaluation}
\noindent\textbf{Quantitative Comparison.} For quantitative comparison, we choose the PSNR and SSIM as metrics, where the higher PSNR and SSIM scores represent the better reconstruction performance. To describe
the performance of each method in detail, we report the PSNR and SSIM scores of different modalities for each client and the average values. It is worth noting that there is a tiny distinction in terms of evaluating different methods. For instance, the \textit{Mixup} method trains a reconstruction model by mixing different modal data, thus, we use the well-trained model to evaluate the performance on the T1w and T2w modalities simultaneously. For the \textit{Group} method, we can obtain two reconstruction models that are trained from different modalities. Then, we evaluate the models on the corresponding modality and record the results respectively. As for the multi-modal reconstruction method (\emph{i.e.,}, \textit{Ideal}, \textit{Gather} and Fed-PMG),  we assign T1w and T2w as an auxiliary modality to reconstruct under-sampled T2w modality and T1w modality, respectively. Here, we only explore the effectiveness of our methods without paying attention to the situation in the real application. Moreover, to make the results more convincing, we evaluate all methods on two different 1D under-sampled masks:  \textit{Random} 4$\times$ and \textit{Equispaced} 4$\times$.

\begin{table*}[!t]
	\centering
		\caption{\small Quantitative comparison of all methods on the fastMRI~\cite{zbontar2018fastmri} dataset with more clients, where \textit{UB} indicates the upper-bound.  The increments and decrements are calculated with the \textit{Group}, and the best and second-best results of FL methods are marked in red and blue, respectively.}
        \small
		\setlength\tabcolsep{2pt}
		\resizebox{1.0\textwidth}{!}{
		\renewcommand\arraystretch{1.5}{
		\begin{tabular}{r||cccc|cccc|cccc}
        \hline
		\multirow{3}{*}{Method}
		&\multicolumn{4}{c|}{\textbf{fastMRI 1.5T}}
		&\multicolumn{4}{c|}{\textbf{fastMRI 3T}}
		&\multicolumn{4}{c}{\textbf{Average }}\\
		\cline{2-13}
		&\multicolumn{2}{c}{\textbf{T1w}}
		
		 &\multicolumn{2}{c|}{\textbf{T2w}}
		 &\multicolumn{2}{c}{\textbf{T1w}}
		
		 &\multicolumn{2}{c|}{\textbf{T2w}}
		 &\multicolumn{2}{c}{\textbf{T1w}}
		
		 &\multicolumn{2}{c}{\textbf{T2w}}\\
		 &PSNR&SSIM&PSNR&SSIM
		 &PSNR&SSIM&PSNR&SSIM
		 &PSNR&SSIM&PSNR&SSIM\\
		 \hline
		  \hline
		  &\multicolumn{12}{c}{\textbf{Random 4$\times$}}\\
		  \hline
		  \textit{Group} & $34.5_{\color{coldgrey}+0.0}$ & $0.923_{\color{coldgrey}+0.000}$ & $32.0_{\color{coldgrey}+0.0}$ & $0.925_{\color{coldgrey}+0.000}$
		  &$34.4_{\color{coldgrey}+0.0}$ & $0.921_{\color{coldgrey}+0.000}$ &$32.5_{\color{coldgrey}+0.0}$ & $0.929_{\color{coldgrey}+0.000}$
		  &$34.4_{\color{coldgrey}+0.0}$ & $0.922_{\color{coldgrey}+0.000}$ &$32.2_{\color{coldgrey}+0.0}$ & $0.927_{\color{coldgrey}+0.000}$ \\
		  
		  \textit{Mixup}&$31.8_{\color{coldgrey}-2.7}$&$0.589_{\color{coldgrey}-0.334}$&$29.6_{\color{coldgrey}-2.4}$&$0.512_{\color{coldgrey}-0.413}$
		  &$31.2_{\color{coldgrey}-3.2}$&$0.496_{\color{coldgrey}-0.425}$&$30.1_{\color{coldgrey}-2.4}$&$0.552_{\color{coldgrey}-0.377}$
		  &$31.5_{\color{coldgrey}-2.9}$&$0.542_{\color{coldgrey}-0.380}$&$29.9_{\color{coldgrey}-2.3}$&$0.532_{\color{coldgrey}-0.395}$ \\
		  
		 \textit{Ideal} & \color{red}{$35.8_{\color{coldgrey}+1.3}$} & \color{blue}{$0.948_{\color{coldgrey}+0.025}$} & \color{blue}{$32.5_{\color{coldgrey}+0.5}$}&\color{blue}{$0.935_{\color{coldgrey}+0.010}$} 
		 &\color{blue}{$35.3_{\color{coldgrey}+0.9}$}&\color{red}{$0.951_{\color{coldgrey}+0.030}$} & \color{blue}{$33.1_{\color{coldgrey}+0.6}$}&\color{red}{$0.942_{\color{coldgrey}+0.013}$}
		 &\color{red}{$35.5_{\color{coldgrey}+0.9}$}&\color{red}{$0.949_{\color{coldgrey}+0.027}$} & \color{blue}{$32.8_{\color{coldgrey}+0.6}$}&\color{blue}{$0.939_{\color{coldgrey}+0.012}$}
		 \\
		 
		 \textit{Gather(UB)}
		 & $36.8_{\color{coldgrey}+2.3}$& $0.955_{\color{coldgrey}+0.032}$ & $33.4_{\color{coldgrey}+1.4}$ & $0.941_{\color{coldgrey}+0.016}$
		 & $37.2_{\color{coldgrey}+2.8}$ &$0.961_{\color{coldgrey}+0.040}$ & $34.0_{\color{coldgrey}+1.5}$ & $0.946_{\color{coldgrey}+0.017}$  
		 & $37.0_{\color{coldgrey}+2.6}$ &$0.958_{\color{coldgrey}+0.036}$ & $33.7_{\color{coldgrey}+1.5}$ & $0.944_{\color{coldgrey}+0.017}$ \\
		 \textbf{Fed-PMG}
		 & \color{blue}{$35.3_{\color{coldgrey}+0.8}$}&\color{red}{$0.950_{\color{coldgrey}+0.027}$}&\color{red}{$32.7_{\color{coldgrey}+0.7}$}&\color{red}{$0.936_{\color{coldgrey}+0.011}$}
		 &\color{red}{$35.8_{\color{coldgrey}+1.4}$}&\color{blue}{$0.946_{\color{coldgrey}+0.025}$}&\color{red}{$33.2_{\color{coldgrey}+0.7}$}&\color{blue}{$0.941_{\color{coldgrey}+0.012}$} 
		 & \color{blue}{$35.5_{\color{coldgrey}+1.1}$} & \color{blue}{$0.948_{\color{coldgrey}+0.026}$} &\color{red}{$32.9_{\color{coldgrey}+0.7}$} &\color{red}{$0.939_{\color{coldgrey}+0.017}$} \\

		  \hline
		  \hline
		  &\multicolumn{12}{c}{\textbf{Equispaced 4$\times$}}\\
		  \hline
		  
		  \textit{Group} & $35.5_{\color{coldgrey}+0.0}$ & $0.930_{\color{coldgrey}+0.000}$ & $33.0_{\color{coldgrey}+0.0}$ & $0.934_{\color{coldgrey}+0.000}$ &$36.8_{\color{coldgrey}+0.0}$ & $0.949_{\color{coldgrey}+0.000}$ &$33.3_{\color{coldgrey}+0.0}$ & $0.930_{\color{coldgrey}+0.000}$
		  &$36.2_{\color{coldgrey}+0.0}$ & $0.940_{\color{coldgrey}+0.000}$ &$33.2_{\color{coldgrey}+0.0}$ & $0.934_{\color{coldgrey}+0.000}$ \\
		  
		  \textit{Mixup}&$32.6_{\color{coldgrey}-2.9}$&$0.584_{\color{coldgrey}-0.346}$&$30.4_{\color{coldgrey}-2.6}$&$0.511_{\color{coldgrey}-0.423}$
		  &$32.0_{\color{coldgrey}-4.8}$&$0.495_{\color{coldgrey}-0.454}$&$31.0_{\color{coldgrey}-2.3}$&$0.564_{\color{coldgrey}-0.366}$
		  &$32.3_{\color{coldgrey}-3.9}$&$0.539_{\color{coldgrey}-0.401}$&$30.7_{\color{coldgrey}-2.5}$&$0.568_{\color{coldgrey}-0.366}$ \\
		  
		 \textit{Ideal} & \color{blue}{$37.3_{\color{coldgrey}+1.8}$} & \color{blue}{$0.953_{\color{coldgrey}+0.023}$} & \color{red}{$34.3_{\color{coldgrey}+1.3}$}&\color{red}{$0.947_{\color{coldgrey}+0.013}$} 
		 &\color{blue}{$37.4_{\color{coldgrey}+0.6}$}&\color{red}{$0.953_{\color{coldgrey}+0.004}$} & \color{blue}{$33.9_{\color{coldgrey}+0.6}$}&\color{blue}{$0.944_{\color{coldgrey}+0.014}$}
		 &\color{blue}{$37.3_{\color{coldgrey}+1.1}$}&\color{blue}{$0.952_{\color{coldgrey}+0.012}$} & \color{red}{$34.6_{\color{coldgrey}+1.4}$}&\color{red}{$0.950_{\color{coldgrey}+0.016}$}
		 \\
		 \textit{Gather(UB)}
		 & $38.4_{\color{coldgrey}+2.9}$& $0.961_{\color{coldgrey}+0.031}$ & $34.4_{\color{coldgrey}+1.4}$ & $0.949_{\color{coldgrey}+0.016}$ 
		 & $38.7_{\color{coldgrey}+1.9}$ &$0.968_{\color{coldgrey}+0.019}$ & $35.5_{\color{coldgrey}+2.2}$ & $0.955_{\color{coldgrey}+0.025}$
		 & $38.5_{\color{coldgrey}+1.1}$ &$0.965_{\color{coldgrey}+0.025}$& $35.1_{\color{coldgrey}+1.9}$ & $0.952_{\color{coldgrey}+0.018}$ \\
		 \textbf{Fed-PMG} 
		 & \color{red}{$37.4_{\color{coldgrey}+1.9}$}&\color{red}{$0.954_{\color{coldgrey}+0.024}$}&\color{blue}{$34.0_{\color{coldgrey}+1.0}$}&\color{blue}{$0.945_{\color{coldgrey}+0.011}$}
		 &\color{red}{$37.4_{\color{coldgrey}+0.6}$}&\color{blue}{$0.953_{\color{coldgrey}+0.004}$}&\color{red}{$34.7_{\color{coldgrey}+1.4}$}&\color{red}{$0.950_{\color{coldgrey}+0.020}$} 
		 & \color{red}{$37.4_{\color{coldgrey}+1.2}$} & \color{red}{$0.954_{\color{coldgrey}+0.014}$} &\color{blue}{$34.4_{\color{coldgrey}+1.2}$} &\color{blue}{$0.947_{\color{coldgrey}+0.013}$} \\
		\hline
		\end{tabular}
		}}
	\label{tab2}
\end{table*}

We report the results of all methods in Table~\ref{tab1}. As we can see, due to utilizing the complementary information of paired multiple modalities, the multi-modal methods significantly outperform the single-modal methods on two modalities over two masks, which improves the average PSNR scores from $32.0$ dB to $35.0$ ($3.0$ dB) on T1w modality and from $31.8$ dB to $33.1$ dB ($1.3$ dB) on T2w modality in terms of \textit{Random 4$\times$} mask, specifically. Besides, the PSNR and SSIM scores of each client in the multi-modal methods are significantly higher than the client from single-modal methods. The results have demonstrated the superiority of multi-modal methods. However, due to the domain shift from different clients, there is a gap between the \textit{Ideal} and the \textit{Gather}, the theoretical upper
bound of multi-modal FL methods. Those results are consistent with the previous work~\cite{guo2021multi,feng2022specificity}, which indicates the performance of FedAvg will degenerate on heterogeneous data. Sharing the distribution information of clients can solve the above problems to some extent~\cite{liu2021feddg,jiang2022harmofl}.
Therefore, the reconstruction performance of our method is close to the \textit{Ideal}, a complete multi-modal federated learning scenario, which indicates that the federated pseudo modality generation scheme can effectively complement the missing modality of the client and learn a good multi-modal reconstruction model.

\noindent\textbf{Qualitative Comparison.}  To qualitatively evaluate our method, we show the restored MR images and corresponding error maps of all methods for two modalities in Fig.~\ref{tab:errormap}. For a comprehensive comparison, we present the results of each client. The less texture in error maps represents the better reconstruction performance.  As we can see, compared with the single-modal methods, the multi-modal methods have better reconstruction performance, which indicates that fusing multi-modal information can significantly improve the reconstruction quality. Besides, the reconstructed images of our method are closed to the \textit{Ideal} method and even better in some cases, which demonstrated the effectiveness of our Fed-PMG for IMFL, \emph{i.e.,} our Fed-PMG can help the incomplete client to train a multi-modal model.

\begin{figure}[!t]
    \centering
    \includegraphics[width=1.0\linewidth]{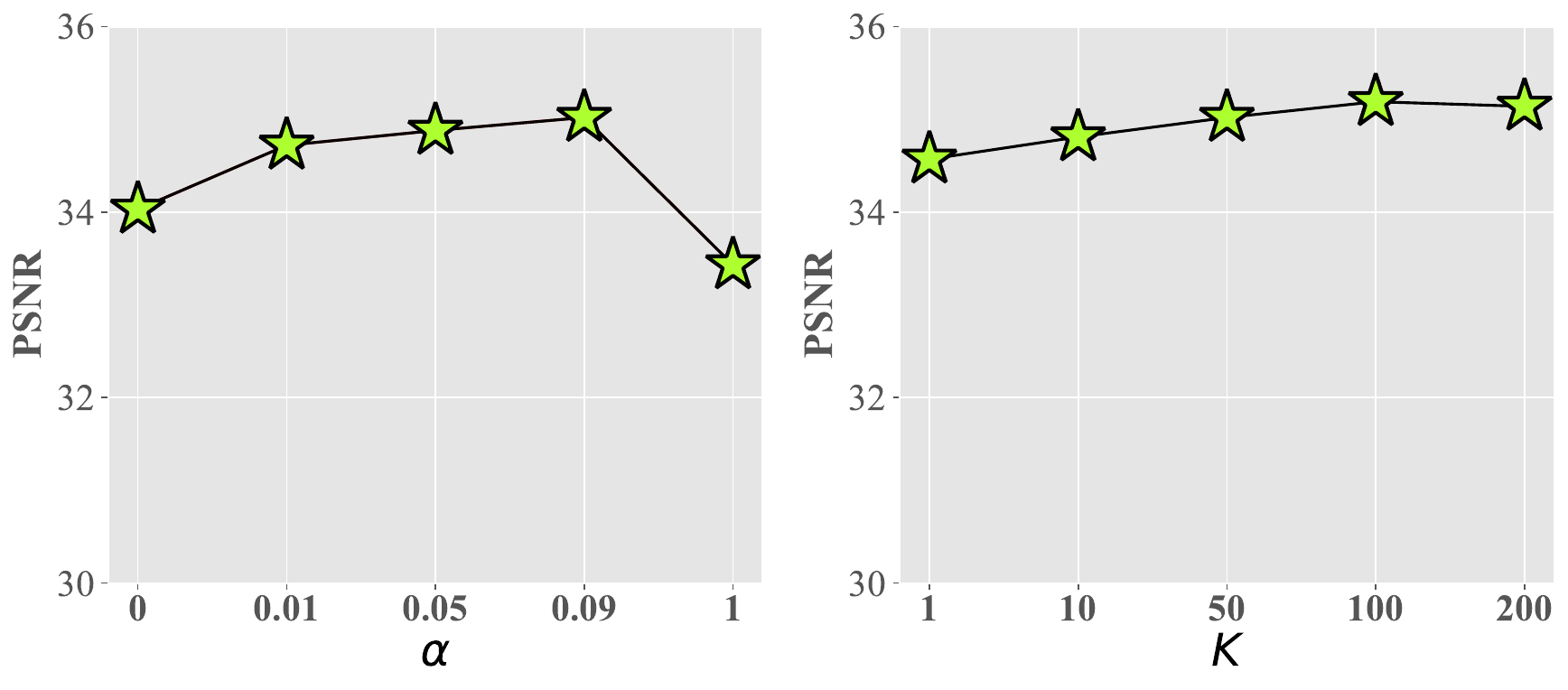}
    \caption{Analysis of hyperparameters $\alpha$ and K in terms of average PSNR on T1w of three clients.} \label{fig:para}
\end{figure}

\subsection{Scalability}
To explore the scalability of our method, we build up a larger federation on fastMRI 3T and fastMRI 1.5T datasets. Specifically, we partition the trainset of each dataset into three partitions. The three clients from the same dataset have different modalities, respectively (\emph{i.e.,} T1w, T2w, T1w and T2w). Thus, there are six clients in total where two clients are multi-modal clients and four clients are single-modal clients. We evaluate all methods on the above new client setting while keeping the other settings default. Table~\ref{tab2} shows the experiment results on the two different masks. From the results, we can observe that our method is still effective even in a larger federation, \emph{i.e.,} our method achieves the similar performance with the \textit{Ideal} and has the higher PSNR and SSIM scores than single-modal methods. The experimental results demonstrate our method is scalable and can handle different numbers of clients.  

\subsection{Hyperparameter Analysis}
To explore the two key hyperparameters, \emph{i.e.,} the control parameter $\alpha$ and the number of cluster $K$, we tune $\alpha$  from candidate set \{0, 0.01, 0.05, 0.09, 1\} and  $K$ from the candidate set \{1, 10, 50, 100, 200\}, respectively, and record the PSNR scores. The above experiments are conducted on three clients, \emph{i.e.,} fastMRI 1.5T, fastMRI 3T, and uMR, with \textit{Equispaced 4×} mask. And the $\alpha$ and $K$ are controlled as $0.09$ and $K=50$ when we tune $K$ and $\alpha$, respectively. 

As shown in Fig.~\ref{fig:para}, the small $\alpha$ is better, while the larger $\alpha$ can narrow the gap of distribution between the pseudo modality and the real modality but increase the variations (artifacts)~\cite{yang2020fda}. Thus, the PSNR score greatly drops when $\alpha=1$, a completely artificial modality that deviates from the real data. Besides, we can observe that the PSNR score is significantly decreased when we set $\alpha=0$. Because when $\alpha=0$, the single-modal clients reconstruct the under-sampled T1w images with self-guide during the training stage. There is a domain gap between the T1w and T2w which introduces the difference in input between training and testing. In terms of $K$, the cluster centroid of the amplitude spectrum can capture more distribution information with the larger $K$. Thus, the generated pseudo modalities can have more chances to cover the missing modalities. However, when $K$ is large enough, increasing $K$ will not improve the performance means that there exists redundant information in the set of the amplitude spectrum. To balance the communication cost and performance, we set $K=50$ as the default in this paper.

\begin{figure}[!t]
    \centering
    \includegraphics[width=1.0\linewidth]{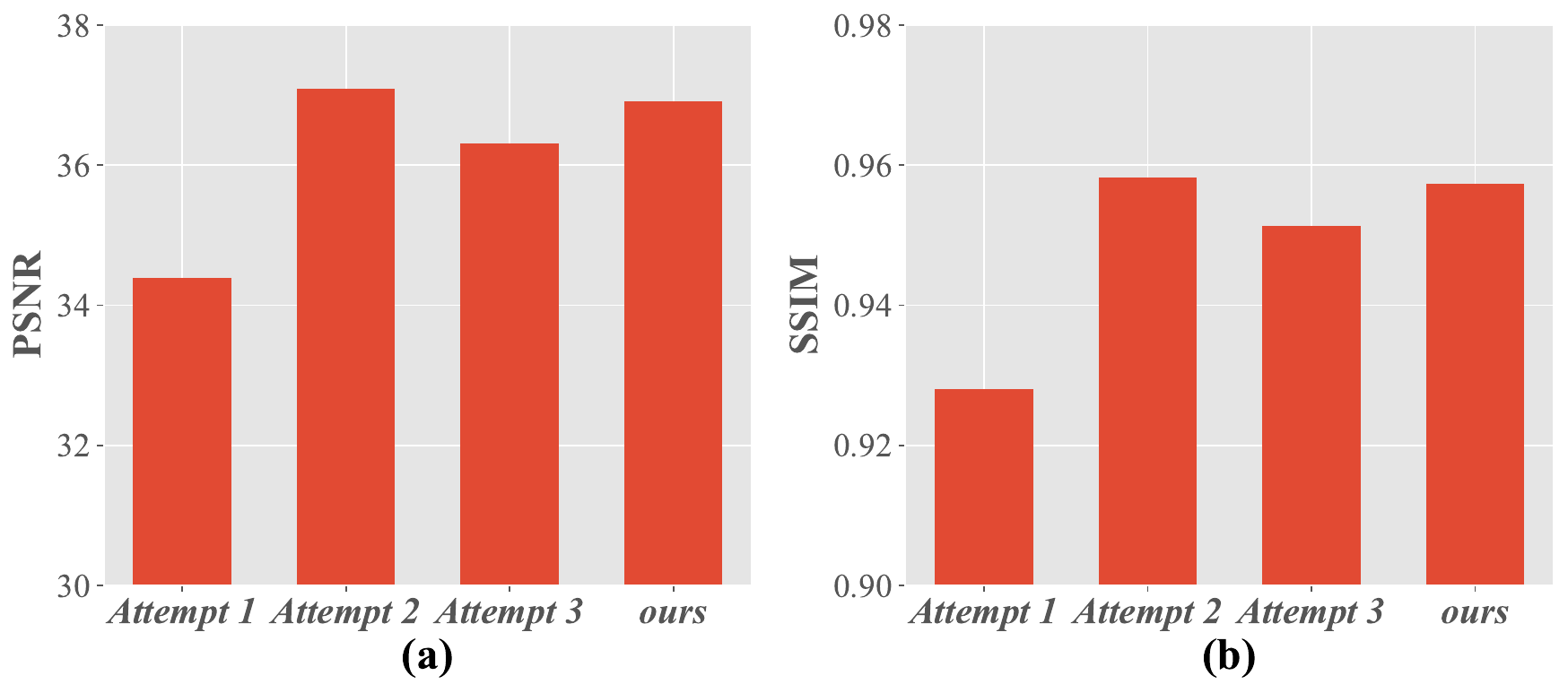}
    \caption{Analysis of sharing information in terms of average PSNR and SSIM on T1w of three clients.} \label{fig:bar}
\end{figure}
\begin{table}[!t]
\centering
\caption{The communication cost of different methods at the fastMRI 1.5T client. \textbf{\#Param.cost}: the cost of sharing model parameters , \textbf{\#Info.cost}: the cost of sharing information. }
\small
\resizebox{0.48\textwidth}{!}{
\renewcommand\arraystretch{1.5}{
\begin{tabular}{lccc}
\toprule
Method  & \#Param.cost (M) & \#Info.cost (M) & Complexity \\
\midrule
\textit{Ideal} & 7.40 & 0 & $\Omega(\varphi)$ \\
Fed-PMG$^\dagger$ & 7.40 & 2227.34 & $\Omega(\varphi + n \beta)$ \\
Fed-PMG & 7.40 & 46.88 & $\Omega(\varphi + K \beta)$ \\

\bottomrule
\end{tabular}}}
\label{tab:cost}
\end{table}

\subsection{Communication Cost}
\label{sec:cost}
We further explore the communication cost before and after the clustering scheme. Therefore, we show the communication cost of one round at the fastMRI 1.5 T client with default setting in Table~\ref{tab:cost}. We also present the space complexity of different methods, where $\varphi$ represents the size of model parameters, $n$ represents the number of images, $\beta$ represents the size of the image, and Fed-PMG$^\dagger$ represents removing the clustering scheme from our method. 
 
As we can see, the cost difference between the three methods is focused on the information cost. Specifically, the cost of shared information is linearly growing as increasing the data quantity when directly sharing all amplitude spectrum, \emph{i.e.,} Fed-PMG$^\dagger$. In contrast, our method can decouple the data quantity and shared information cost, thus the space complexity of our method is constant. Compared with Fed-PMG$^\dagger$, our method reduces the cost by about $97.5$\%. Due to the use of clustering strategy, the cost of our method is acceptable. 

 \begin{table}[!t]
    \centering
        \caption{\textbf{Quantitative comparison of all competing methods} under a higher domain shift. We utilize different undersample patterns for three clients: \textit{Equispaced} 4$\times$, \textit{Random} 6$\times$, and \textit{Random} 8$\times$. }
    \label{gb2}
    \small
    \resizebox{0.48\textwidth}{!}{\setlength{\tabcolsep}{7pt}{
    \renewcommand\arraystretch{1.3}{
    {\begin{tabular}{l|cc|cc}
    \hline
        \multirow{3}{*}{Method} &\multicolumn{4}{c}{\textbf{Average}} \\
    \cline{2-5}  & \multicolumn{2}{c|}{T1w} & \multicolumn{2}{c}{T2w} \\
    & PSNR & SSIM &PSNR &SSIM  \\
       \hline
    \textit{Mixup} & 33.0 & 0.916 & 30.2 & 0.839 \\ 
    \textit{Group} & 33.8 & 0.937 & 31.0 & 0.887 \\
     \textit{Ideal} & 34.3 & 0.941 & 32.2 & 0.934  \\
     \textbf{Fed-PMG (ours)} & \textbf{34.6} & \textbf{0.951} & \textbf{32.1} & \textbf{0.935} \\
    \hline
    \end{tabular}}}}}
\end{table}

\subsection{Ablation Study} \label{sec:abl}

\noindent{\textbf{Analysis of Shared Amplitude centroid.}} To further investigate the proposed federated pseudo modality generation scheme, we have made some attempts to analyse the effectiveness of shared amplitude centroid. To be specific, we compared various pseudo modality generation methods, which employ different information as follows:

\begin{itemize}
	\setlength{\itemsep}{3pt}
	\setlength{\parsep}{-2pt}
	\setlength{\parskip}{-0pt}
	\setlength{\leftmargin}{-15pt}
	\item \noindent{\textbf{\textit{Attempt 1: Generating pseudo modality with all-zeros matrix.}}} Since the phase spectrum can capture the structure information of images, which is crucial for multi-modal reconstruction, a naive idea is to use an all-zeros matrix as an amplitude spectrum and combine it with the phase spectrum to generate a pseudo modality.
    \item \noindent{\textbf{\textit{Attempt 2: Generating pseudo modality with the shared amplitude spectrum.}}}  We directly share the amplitude spectrum memory among different clients instead of the cluster centroid of amplitude spectrum. Thus, the difference with our method is removing the clustering scheme.

    \item \noindent{\textbf{\textit{Attempt 3: Generating pseudo modality with intra-client amplitude spectrum.}}} We randomly select the amplitude spectrum of unused modality and combine it with the phase spectrum to generate pseudo modality in the single-modal client. The difference with \textit{Attempt 2} is that it only uses the intra-client amplitude spectrum instead of shared amplitude spectrum memory that contains the distribution information of other clients.
    
\end{itemize}
We conduct the above attempts and our method on the T1w modality of three clients with \textit{Equispaced 4×} under-sampled mask. The average PSNR and SSIM scores of all methods are presented in Fig.~\ref{fig:bar}. As we can see, the performance of \textit{Attempt 1} is significantly lower than other methods, which indicates that the amplitude spectrum is also important for the reconstruction. Besides, comparing \textit{Attempt 2} and \textit{Attempt 3}, the results demonstrate that sharing distribution information of amplitude spectrum can mitigate the domain shift to some extent. More importantly, the performance of our method is close to the \textit{Attempt 2} that directly shares the amplitude spectrum. And the previous experiment results show that our method can reduce a lot of communication loss compared with \textit{Attempt 2} (see \S\ref{sec:cost}). Hence, our approach achieves a balance between communication cost and performance by clustering scheme.

\noindent\textbf{Robustness toward Domain Shift}. 
To explore the robustness of our method towards domain shift, we conduct a comparison with a higher domain shift by using different undersample patterns~\cite{feng2022specificity}, \textit{i.e.,} \textit{Equispaced} 4$\times$, \textit{Random} 6$\times$, and \textit{Random} 8$\times$ at three clients for different methods. As shown in Table~\ref{gb2}, our method also outperforms all single-modal FL methods in this larger domain shift setting and performs similarly to \textit{Ideal}. The results indicate that our method is robust for different heterogeneous settings.

\begin{table}[!t]
    \centering
        \caption{\textbf{Quantitative comparison of all competing methods} with different backbones on three clients under Equispaced 4$\times$.}
    \small
    \resizebox{0.42\textwidth}{!}{\setlength{\tabcolsep}{4pt}{
    \renewcommand\arraystretch{1.3}{
    {\begin{tabular}{l|c|cc|cc}
    \hline
    \multirow{3}{*}{Method}&\multirow{3}{*}{Backbone}
    &\multicolumn{4}{c}{\textbf{Average}} \\
    \cline{3-6}
    && \multicolumn{2}{c|}{T1w} & \multicolumn{2}{c}{T2w} \\
    & & PSNR & SSIM &PSNR &SSIM  \\
       \hline
    \textit{Mixup} & \multirow{4}{*}{Unet} & 31.6 & 0.776 & 31.0 & 0.753 \\ 
    \textit{Group} & & 35.4 & 0.939 & 33.2 & 0.934 \\
     \textit{Ideal} & & 36.8 & 0.959 & 34.5 & 0.950  \\
     \textbf{Fed-PMG (ours)} & & \textbf{36.9} & \textbf{0.957} & \textbf{34.6} & \textbf{0.951} \\
    \hline
\textit{Mixup}&\multirow{4}{*}{VN} & 33.5 & 0.883 & 31.8 & 0.811\\ 
     \textit{Group}& & 36.6 & 0.955 & 34.1 & 0.950 \\
     \textit{Ideal}& & 38.0 & 0.971 & 36.2 & 0.970 \\
     \textbf{Fed-PMG (ours)} & & \textbf{38.2} & \textbf{0.973} & \textbf{36.0} & \textbf{0.968} \\
    \hline
    \end{tabular}}}}}
    \label{backbone}
\end{table}

\begin{table}[!t]
    \centering
        \caption{\textbf{Quantitative comparison of all competing methods} under a modality fully missing setting. The
modality of three clients is T1, T2, and T2, respectively}
    \label{fullymissing}
    \small
    \resizebox{0.45\textwidth}{!}{\setlength{\tabcolsep}{7pt}{
    \renewcommand\arraystretch{1.3}{
    {\begin{tabular}{l|cc|cc}
    \hline
        \multirow{3}{*}{Method} &\multicolumn{4}{c}{\textbf{Average}} \\
    \cline{2-5}  & \multicolumn{2}{c|}{T1w} & \multicolumn{2}{c}{T2w} \\
    & PSNR & SSIM &PSNR &SSIM  \\
       \hline
    \textit{Mixup} & 33.0 & 0.916 & 30.2 & 0.839 \\ 
    \textit{Group} & 33.8 & 0.937 & 31.0 & 0.887 \\
     \textit{Ideal} & 34.3 & 0.941 & 32.2 & 0.934  \\
     \textbf{Fed-PMG (ours)} & \textbf{34.6} & \textbf{0.951} & \textbf{32.1} & \textbf{0.935} \\
    \hline
    \end{tabular}}}}}
\end{table}

\noindent\textbf{Different BackBones}. We explore the sensitivity of our method toward different backbones. Specifically, we utilize a new backbone, \textit{i.e.,} VN~\cite{hammernik2018learning}, for all methods on three default clients with \textit{Equispaced} 4$\times$ undersample pattern, as shown in Table~\ref{backbone}. Apparently, although the performance of all methods is improved with a strong backbone (VN), our Fed-PMG also outperforms all competitors, which shows the improvement over other methods is robust for different backbones.

\noindent\textbf{Performance under Modality Fully Missing Setting}. The sole assumption underlying this method is the presence of all modalities within the federation, thus making it unrestricted to multi-modal clients. To validate this claim, we establish a new modality fully missing setting where the modalities of three clients are T1, T2, and T2, respectively. As shown in Table~\ref{fullymissing}. The results indicate our method still outperforms other FL methods.

\subsection{Combining with Heterogeneous FL Methods}
The primary goal of our method is to address the incomplete multi-modal data in federated learning, though we have confirmed it can mitigate domain shift to a certain extent in \S\ref{sec:abl} by sharing the cross-client centroid, which contains the distribution information of the client. Therefore, our method is scalable and can combine with other FL approaches to address the domain shift. To explore this, we combine our method with several representative heterogeneous FL methods:
\begin{itemize}
	\setlength{\itemsep}{3pt}
	\setlength{\parsep}{-2pt}
	\setlength{\parskip}{-0pt}
	\setlength{\leftmargin}{-15pt}
    \item \textbf{FedBN}~\cite{li2021fedbn}: It is a general solution heterogeneous FL setting, which can address the domain shift by client-specific Batch Normalization layer.
    \item \textbf{FL-MRCM}~\cite{guo2021multi}: This is the first heterogeneous FL method for MRI reconstruction, which utilizes adversarial learning to align the features between source and target sites.
    \item \textbf{FedMRI}~\cite{feng2022specificity}: It introduced the personalized FL to address the domain shift among different clients. 
\end{itemize}

The primary goal of these two methods is to solve the non-IID data issue in FL, while our method is to address the incomplete multi-modal MRI FL issue. That is, our method can be easily plugged into them. We present the results of these heterogeneous FL methods and combination methods in Table~\ref{gp}. As we can see, the above heterogeneous FL methods can effectively mitigate the domain shift, which yields a significant improvement compared with Mixup (\textit{i.e.,} FedAvg). However, their performance is still lower than the multi-modal methods. This indicates the primary problem in this setting is modality missing. In addition, our method can further improve the performance based on FedBN, FL-MRCM, and FedMRI, as they can alleviate the data heterogeneity in FL. 

\begin{table}[!t]
    \centering
    \caption{\textbf{The results of combination method} on three clients under Equispaced 4$\times$.}
    \label{gp}
    \resizebox{0.50\textwidth}{!}{\setlength{\tabcolsep}{6pt}{
    \renewcommand\arraystretch{1.3}{
    \begin{tabular}{l|cc|cc}
    \hline
    \multirow{3}{*}{Method} &\multicolumn{4}{c}{\textbf{Average}} \\
    \cline{2-5}
    &\multicolumn{2}{c|}{T1w} & \multicolumn{2}{c}{T2w} \\ 
    & PSNR & SSIM &PSNR &SSIM \\
       \hline
    \textit{Group} & 35.4 & 0.939 & 33.2 & 0.934 \\
    \textit{Mixup} & 31.6&0.776&31.0&0.753 \\
    \textit{FedBN} & 32.8 & 0.823 & 31.6 & 0.798 \\
    \textit{FL-MRCM} & 31.9 & 0.787 & 31.5 & 0.768 \\
    \textit{FedMRI} & 34.7 & 0.908 & 32.6 & 0.894 \\ 
    \textbf{Fed-PMG} & \textbf{36.9} & \textbf{0.957} & \textbf{34.6} & \textbf{0.951}\\
    \hline
    \textit{Group} + FedBN&  35.7 & 0.941 & 33.3 & 0.935 \\
    \textbf{Fed-PMG + FedBN}&  \textbf{37.1} &\textbf{0.959} & \textbf{34.8} & \textbf{0.953} \\
    \hline
    \textit{Group} + FL-MRCM & 36.2 & 0.948 & 33.5 & 0.939 \\
    \textbf{Fed-PMG + FL-MRCM}&  \textbf{37.4} & \textbf{0.961} & \textbf{35.1} & \textbf{0.955} \\
    \hline
    \textit{Group} + FedMRI & 36.5 & 0.950 & 34.1 & 0.942 \\ 
    \textbf{Fed-PMG  + FedMRI} & \textbf{38.3} & \textbf{0.972}  &\textbf{35.3} & \textbf{0.963} \\
    \hline
    \end{tabular}}}}
\end{table}

\section{Conclusion}
In this paper, we studied a novel multi-modal MRI reconstruction problem in federated learning, which had a more complex situation compared with the single-modal reconstruction task. The main challenge in our FL setting is the unpaired modality, which makes it impossible to train a multi-modal model. To tackle this challenge, we propose a federated pseudo modality generation scheme to generate the missing modality. Meanwhile, our method can effectively mitigate the domain shift due to the distribution information of the shared amplitude spectrum. Extensive experiments demonstrate that our method has the better performance than the single-modal method and even achieves a similar effect with the ideal (modal-complete) scene.

{\small
\bibliographystyle{IEEEtran}
\bibliography{tmi}
}

\end{document}